# On the Accuracy of Galileo's Observations

**(post-publication version with large figures)**

Christopher M. Graney

Jefferson Community College
1000 Community College Drive
Louisville, Kentucky 40272, USA
(502) 213-7292
christopher.graney@kctcs.edu
www.jefferson.kctcs.edu/faculty/graney

**ABSTRACT**

Galileo Galilei had sufficient skill as an observer and instrument builder to be able to measure the positions and apparent sizes of objects seen through his telescopes to an accuracy of 2" or better. However, Galileo had no knowledge of wave optics, so when he was measuring stellar apparent sizes he was producing very accurate measurements of diffraction artifacts and not physical bodies.

## 1. INTRODUCTION

Previous work in this journal [*Baltic Astronomy*] by Standish and Nobili (1997) has illustrated that Galileo's careful observations, measurements, and to-scale drawings of the Jovian system improved in accuracy from their commencement in 1610 to the point that by January 1613 Galileo was recording the separations between Jupiter and its moons to within 0.1 Jovian radii (approximately 2"), placing Jupiter's moons in his drawings to an accuracy of better than the width of the dots he used to mark the moons' positions, and recording in his drawings positions of objects as faint as Neptune (Standish and Nobili 1997). That he did this using a small "Galilean" telescope that lacked even a focal plane in which to place a measuring reticle makes this feat all the more remarkable.

Evidence of Galileo's skill is not limited to these observations of Jupiter. Over a span of two decades he often wrote as though he could regularly achieve accuracies of 2" or better. His notes demonstrate this degree of accuracy in his measuring and drawing the positions and sizes of celestial objects. He was aware of his skill and of the quality of his instruments and had confidence in the repeatability of his data. However, since Galileo lacked understanding of wave optics, when it came to stellar observations, often what he was measuring so accurately was diffraction artifacts.

## 2. GALILEO'S MEASUREMENTS OF POSITION

In 1612 Galileo made an assessment of his improving ability to make accurate measurements – improving ability that Standish and Nobili would later discover. In his "Discourse on Bodies Floating in Water" Galileo reported that he had improved his ability to make measurements in the Jovian system to the point that he could measure to an accuracy of arc seconds, whereas previously he had only been able to achieve an accuracy of an arc minute (*Le Opere di Galileo*, IV, p. 64). An interesting illustration of Galileo's observing skill in this regard can be seen by comparing some of Galileo's drawings of the Jovian system (including the one that Standish and Nobili discovered includes Neptune) to simulated telescopic views generated by planetarium software [Figure 1].

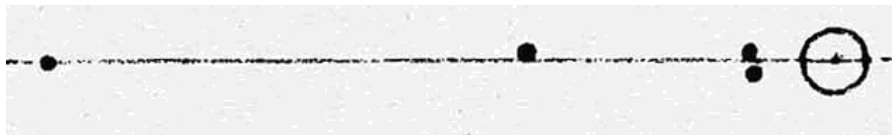
Galileo
25 March 1613 H 0.5
(*Opere* V p. 241)

*Stellarium*
25 March 1613
12:56:00 EST
FOV 0.367°

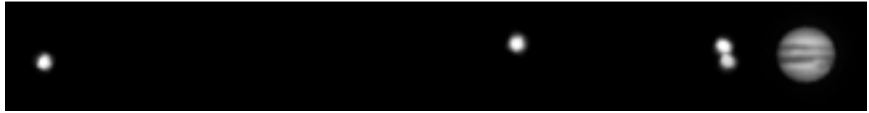
Galileo
12 March 1613 H 5
(*Opere* V p. 241)

*Stellarium*
12 March 1613
4:52:00 PM EST

FOV 0.366°

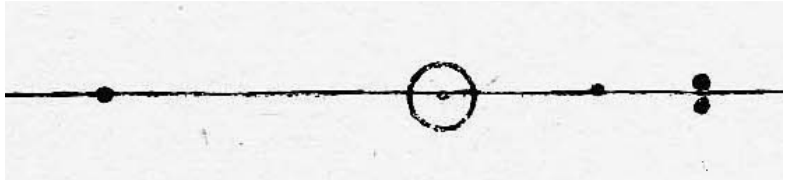
Galileo
29 March 1613
H 0.0.30
(*Opere* V p. 243)

*Stellarium*
29 March 1617
12:52:00 PM EST
FOV 0.368°

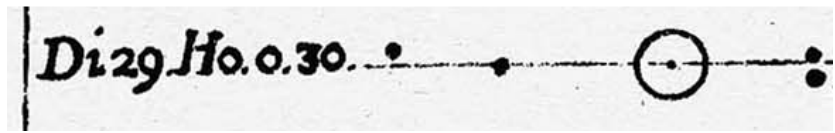
Galileo
6 January 1613
(Standish & Nobili)

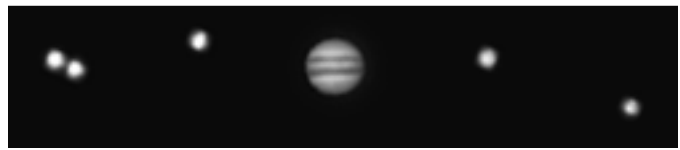
*Stellarium*
6 January 1613
12:02:58 AM EST
FOV 0.368°

**Figure 1:** Galileo's drawings compared to output from the *Stellarium* open-source planetarium software (www.stellarium.org). 6 January 1613 shows Neptune in the lower-right corner. *Stellarium* gives Neptune's magnitude as being 7.9 at the time. The *Stellarium User Guide* states that the positions of Jupiter and Neptune are accurate to 1", and that positions of Galilean satellites are valid for 500 A.D. – 3500 A.D. (no level of accuracy given). Differences exist between drawings and *Stellarium* due not only to Galileo's and *Stellarium's* errors, but also to the author's errors in estimating the precise moment in time that a given drawing represents.

In January of 1617 Galileo observed but did not draw the double star Mizar in Ursa Major and measured its angular separation to be 15″.  A copy of Galileo's original notes on this observation can be found in Ondra (2004); a scholarly discussion of them is available in Seibert (2005); the notes are also available in *Le Opere di Galileo*, III, PT. 2, p. 877.  This 15″ value is within an impressive half an arc second of modern measurements.

In February 1617 Galileo observed and made a drawing of a grouping of stars in Orion in the region of the Trapezium (Seibert 2005; *Le Opere di Galileo*, III, PT. 2, p. 880).  A comparison of that drawing to modern data on those stars illustrates that Galileo's skill and the quality of his instruments were sufficient for him to produce a very accurate record of stars that were separated by less than 15″.  Figure 2 shows Galileo's drawing of five stars in Orion (which Galileo labeled a, b, c, g, i) as printed in *Le Opere di Galileo*; a chart of the locations of stars HD 37042, HD 37041, HD 37023, HD 37022, and HD 37020 from the Trapezium region plotted according to their ICRS 2000.0 coordinates as given by the SIMBAD astronomical database (http://simbad.harvard.edu/Simbad -- none of the stars in question have large enough proper motions to produce substantial changes between 1617 and today); and a superposition of these two, with Galileo's sketch processed to show his markings as white areas circled by a black border, and rotated and enlarged to match the SIMBAD position chart.  Figure 2 also shows the same method applied to a January 2001 European Southern Observatory image of the same region of the sky, for the sake of comparison.

### 3. GALILEO'S MEASUREMENTS OF STELLAR SIZE

Galileo also measured the apparent angular sizes of stars.  These include Sirius, whose apparent diameter he measured to be just over 5″ (*Le Opere di Galileo*, III, PT. 2, p. 878), and the components of Mizar, whose apparent diameters he measured in 1617 as being 6″ and 4″ (Ondra 2004, Siebert 2005, *Le Opere di Galileo*, III, PT. 2, p. 877).  Both of these measurements are from Galileo's notes and were unpublished.

Galileo apparently measured the sizes of other stars as well.  In a 1624 letter to Francesco Ingoli he reports

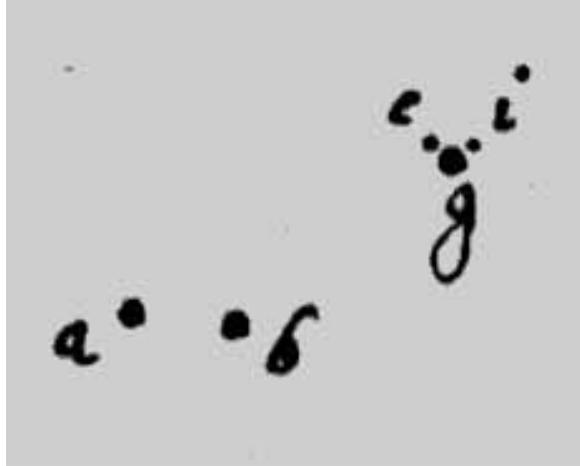

**Figure 2A:**

Top left -- Galileo's drawing of five stars in Orion as printed in *Le Opere di Galileo*.

Middle -- chart of the locations of stars HD 37042, HD 37041, HD 37023, HD 37022, and HD 37020 from the Trapezium region.

Bottom left -- superposition of the two, with Galileo's sketch processed to show his markings as white areas circled by a black border, and rotated and enlarged to match the chart.

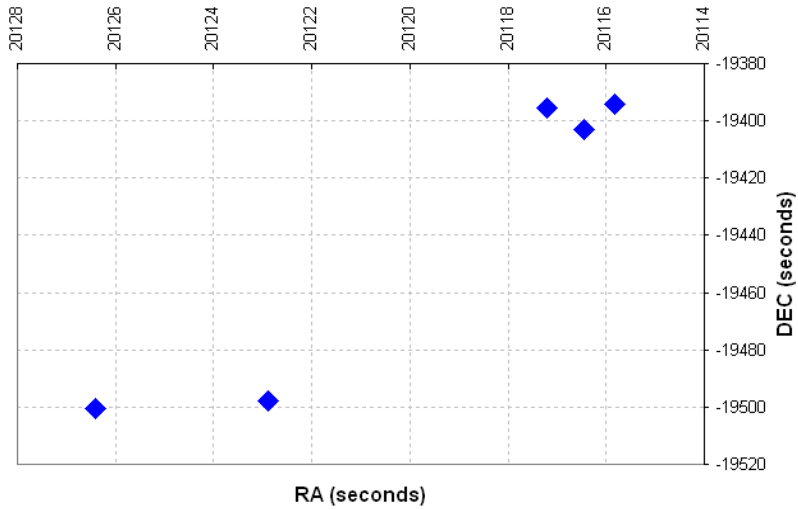

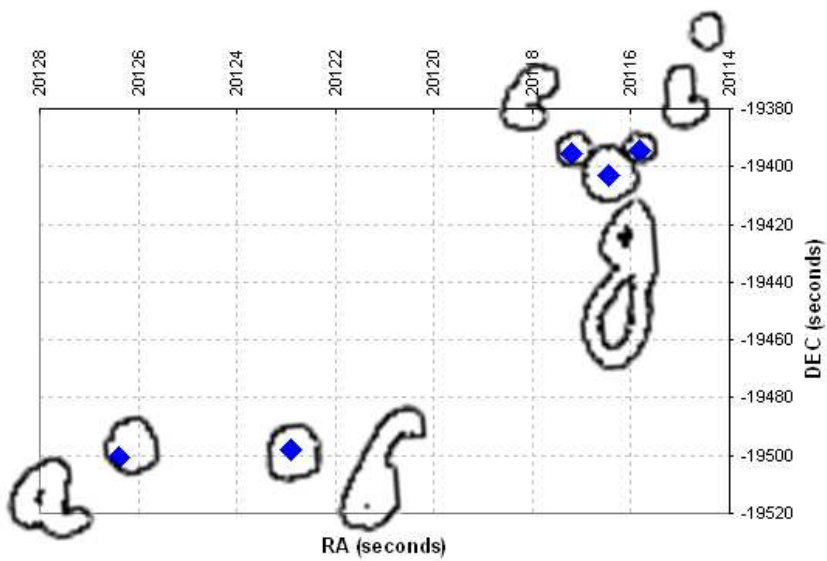

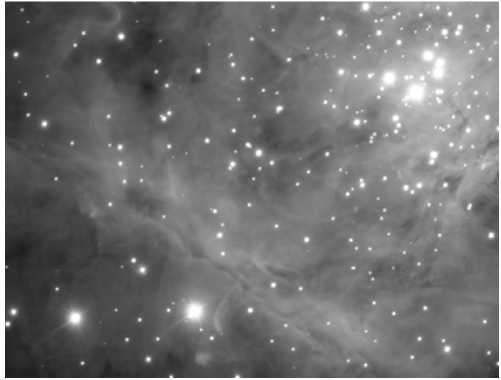

http://www.eso.org/outreach/gallery/vlt/images/
Top20/Top20/press-rel/phot-03a-01-normal.jpg

**Figure 2B:**

The same method (as figure 2A) applied to a January 2001 European Southern Observatory image of the same region of the sky.

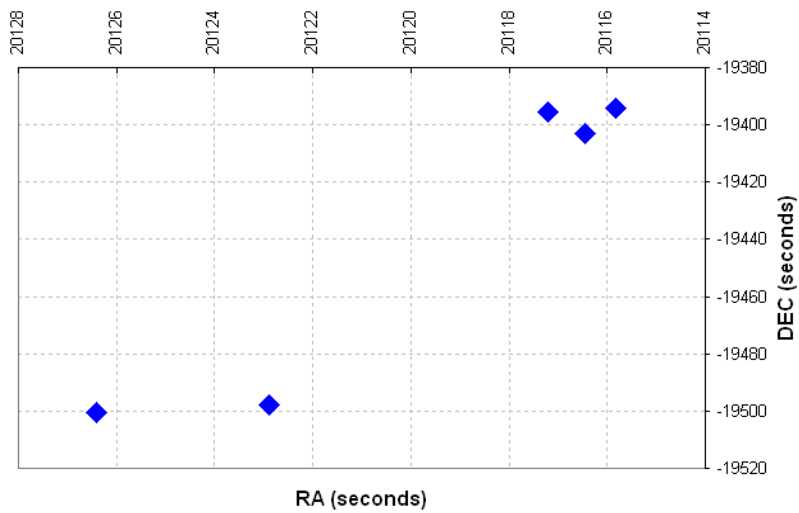

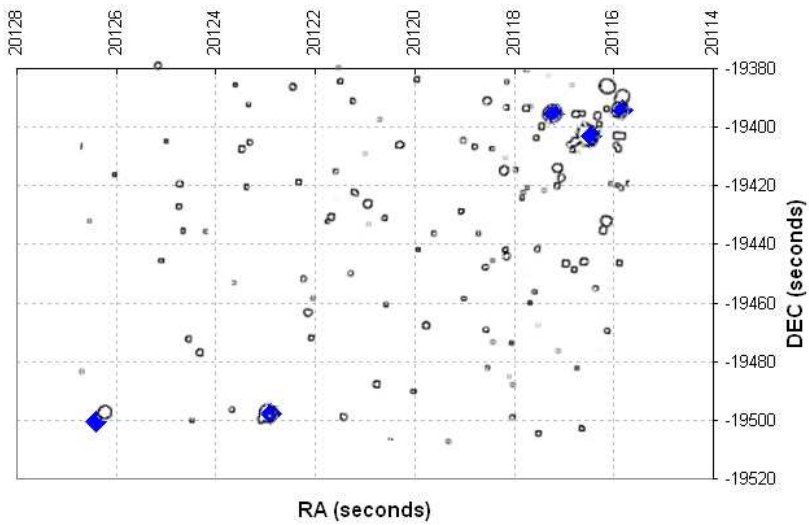

knowing from experience that no star subtends more than 5″, and that a great number subtend less than 2″ (Finocchiaro 1989, p. 174).  In his *Dialogue Concerning the Two Chief World Systems* published eight years later, Galileo states that a first-magnitude star has a diameter of 5″ while a sixth-magnitude star has a diameter of one-sixth that, and implies a linear relationship exists between magnitude and diameter (Drake 1967, p. 359-362).

From the above information, including the 2″ lower value mentioned in the letter to Ingoli and the Mizar measurement which exceeded the 5″ maximum he later asserts, it seems reasonable to state that Galileo understood that he could reliably measure the sizes of objects to an accuracy of at least 2″.

**4. EFFECTS OF DIFFRACTION**

What Galileo did not understand was that in the case of stellar sizes, 2″ probably represented nothing more than a combination of a wave optics diffraction pattern and the limits of the human eye.  The image of a star formed by a telescope is a diffraction pattern consisting of a central maximum (Airy Disk) whose angular radius is given by $r_A = 1.22\lambda/D$.  $\lambda$ is taken in this paper as 550 nm, the center of the visible spectrum.  It would be more than a century and a half after Galileo before astronomers began to understand and investigate the fact that the apparent size of a star was a product of a telescope's aperture (Herschel 1805).  Galileo had no reason to believe that the apparent size of a star was any more spurious than the apparent size of Jupiter.

While the images of all stars have the same $r_A$ even if their magnitudes differ, they do not all have the same apparent size because they do not all have the same intensities [Figure 3].  A telescope system (telescope, eye, and sky conditions) has an intensity threshold below which the eye detects nothing, and above which the eye detects starlight.  As seen in Figure 3, the result of this threshold is that stars of differing magnitude will have differing apparent sizes, with the relationship appearing linear over a limited range of magnitudes.

Modern interferometric tests on optics Galileo used in his telescopes have shown that he was able to obtain

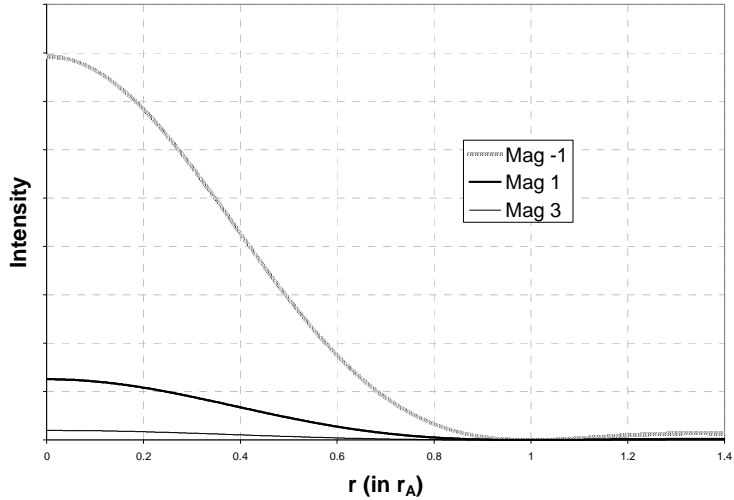
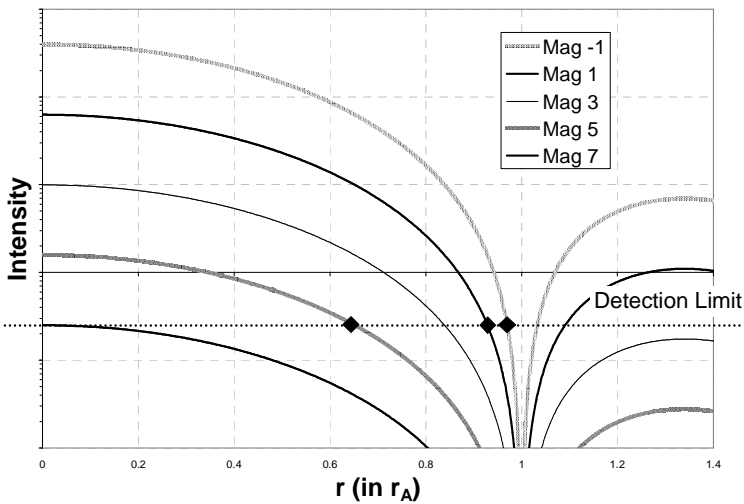
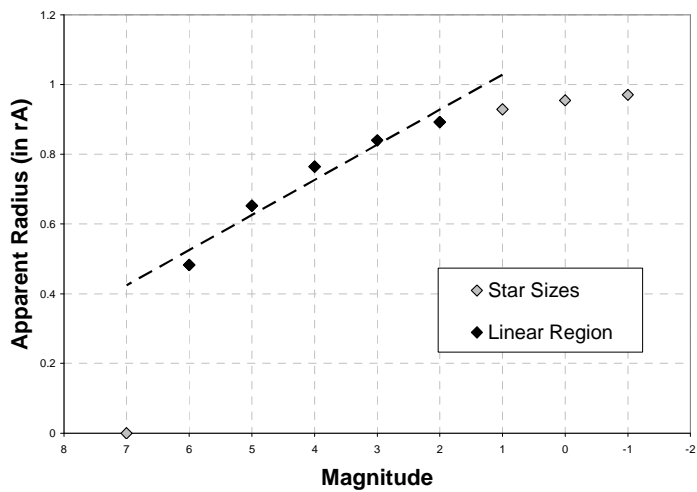

**Figure 3:**

In the diffraction pattern from a circular aperture the intensity as a function of radius is given by $I=I_0[J_1(r)/r]^2$ where $J_1(r)$ is a Bessel function of the first kind.

Top Row -- plot of I vs. r.

Middle Row -- semi-log plot of I vs. r for a system with an intensity threshold such that a star of magnitude 7 cannot be detected (horizontal line). The result of this limit is that the stars will have differing sized apparent radii, shown by the marks on the plot where the stars' intensities drop below the threshold.

Bottom Row -- plot of apparent radius vs. magnitude. Note that for middling magnitude stars, the relationship would appear essentially linear to observers, especially considering that truly bright stars that break from the line are comparably few in number and faint stars that break from the line are a challenge to observe and measure. For a larger telescope, both the limiting magnitude and $r_A$ would be smaller and the linear relationship would be less obvious, while for a smaller telescope the relationship would be more pronounced.

"nearly perfect optical quality" in his instruments; but these optics are small and vary little in aperture (Greco, Molesini, Quercioli 1992). Galileo's telescopes were of a size and quality just right for producing star images of a few arc seconds in diameter on a night of good "seeing". What Galileo thought were the actual sizes of stars were probably artifacts of diffraction.[*]

## 5. DIFFRACTION AND GALILEO'S MEASUREMENTS OF MIZAR'S POSITION AND SIZE

Studying diffraction in the case of his Mizar measurements highlights Galileo's abilities even further. According to the SIMBAD database the magnitudes of Mizar A and B (HD 116656 and HD 116657) are 2.27 and 3.95 respectively and their relative motions are not significant enough to greatly alter their separation between 1617 and today. Their separation according to Hipparcos data from the Millennium star atlas (http://www.rssd.esa.int/Hipparcos/msa-tab7.html) is 14.4″. As mentioned previously, Galileo observed A to have a diameter of 6″ and B to have a diameter of 4″, with a separation of 15″. Telescopes of 26 mm and 38 mm apertures are attributed to Galileo (Greco, Molesini, Quercioli 1992). Assuming these sizes are the result of diffraction, plotting the intensity curves for these two stars based on a 26 mm telescope, and setting a detection threshold such that the image of B will have a 4″ diameter yields an expected diameter for A of 7.3″ [Figure 4], differing from Galileo's measurement by 1.3″. The same calculations performed for a 38 mm telescope also yields results that are not much different from Galileo's measurements [Figure 4]. Regardless of the telescope size used, the agreement between what Galileo observed and the results of the calculations is very close. This reinforces the idea that Galileo could make excellent measurements but that in regards to stellar sizes Galileo was measuring diffraction artifacts.

---

[*] For additional evidence that Galileo's telescopes were of high optical quality and his measurements of stellar diameters are attributable to diffraction, the reader is advised to study Tom Pope and Jim Mosher's web site, "CCD Images from a Galilean Telescope" (www.pacifier.com/~tpope). Pope and Mosher constructed a Galilean telescope and obtained images through it using a CCD camera. By comparing their images with Galileo's notes and sketches, they too find Galileo to be remarkably accurate in his observations.

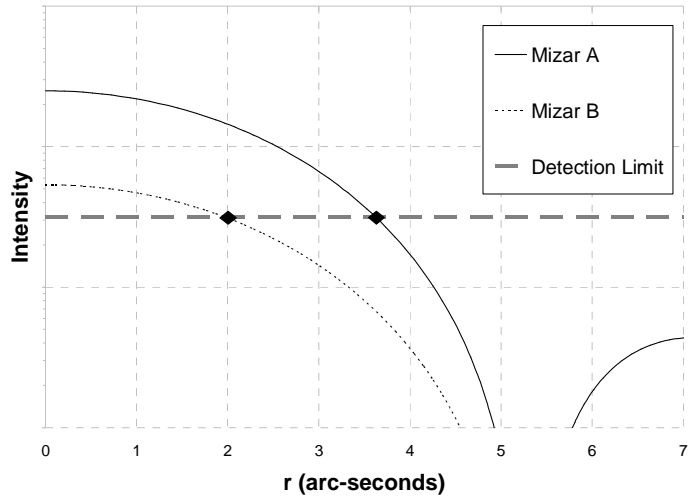
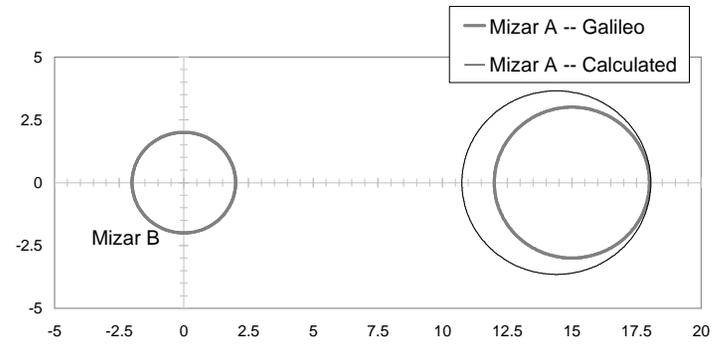
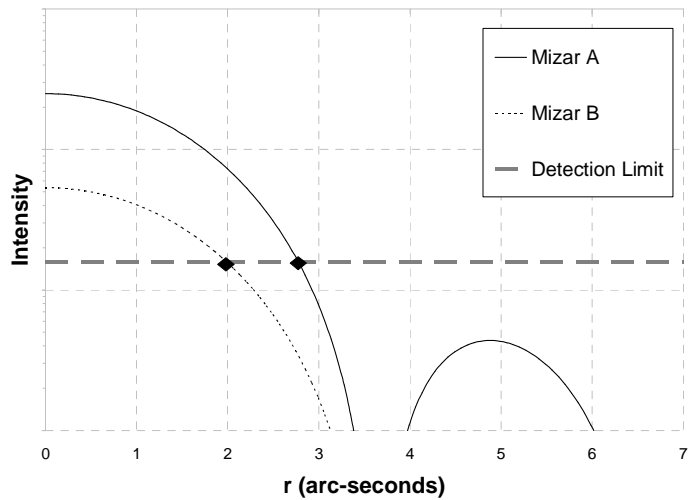
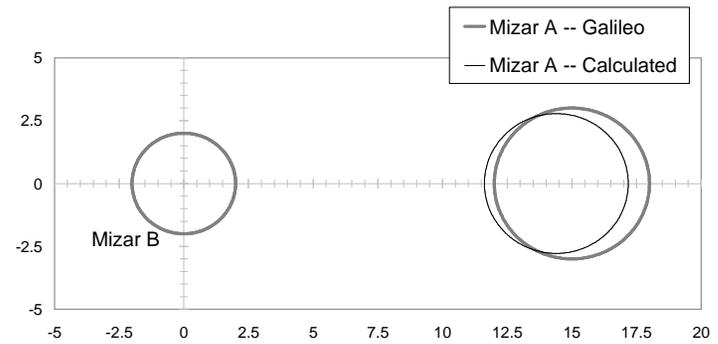

**Figure 4:** Semi-log plots of intensity curves similar to those in Figure 3, but for Mizar A and B based on magnitudes of 2.27 and 3.95 respectively. Detection threshold is set to give B a radius of 2" so as to agree with Galileo's measurement of B. That threshold is then used to determine the apparent radius of A. Top Row -- Plot of the intensity curves for these two stars based on a 26 mm telescope, with a diagram showing Mizar as Galileo measured it and as it would be expected to appear based on intensity curve calculations and a modern value for the separation of the two components. Bottom Row -- Same method applied to a 38 mm telescope instead of a 26 mm telescope.

## 6. CONCLUSIONS

Galileo's work shows that he was capable of achieving an accuracy of 2" or better in measuring and drawing the positions and sizes of celestial objects. Since Galileo was unaware of wave optics, in regards to stellar sizes Galileo was simply measuring artifacts of diffraction. Nonetheless his work shows a remarkable level of skill, and it is clear Galileo was aware of his skill. Taken as a whole, Galileo's measurements of the Jovian system, the Trapezium, Sirius, and Mizar indicate that the accuracy he achieved was not a fluke and the claims he made in his writing were valid. That Galileo, the first scientist to use a telescope to study the heavens, could achieve such results using only his eyes and a telescope that lacked even a basic reticle for measurements is a testament to his talent and work ethic.

ACKNOWLDEGDEMENTS: The author would like to thank Myles Standish for his review of and helpful comments on this paper. The author would also like to thank Edmundas Meištas for his assistance with formatting the text and figures in the published version of this paper.

**REFERENCES**

Drake, S. 1967, *Dialogue Concerning the Two Chief World Systems – Ptolemaic & Copernican, 2$^{nd}$ edition*, Los Angeles: University of California Press.

Finocchiaro, M. A. 1989, *The Galileo Affair – A Documentary History*, Los Angeles: University of California Press.

Galileo Galilei, *Le Opere di Galileo -- Edizione Nazionale Sotto gli Auspicii di Sua Maestà il re d'Italia*, ed. A. Favaro; 20 vols, Florence, 1890-1909.

Greco, V., Molesini, G., Quercioli, F. 1992, Nature, 358, 101.

Herschel, W. 1805, Philosophical Transactions of the Royal Society of London, 95, 31.

Ondra, L. July 2004, Sky and Telescope.

Siebert, H. 2005, Journal for the History of Astronomy, 36, 251.

Standish, N., Nobili, A. 1997, Baltic Astronomy, 6, 97.